\documentclass[aps,prl,groupedaddress,reprint,superscriptaddress,floatfix,tightenlines ]{revtex4-1} 
\usepackage{lipsum}
\usepackage{graphicx,import}
\usepackage{grffile} 
\usepackage[usenames,dvipsnames,svgnames,table]{xcolor}
\setkeys{Gin}{width=\columnwidth} 
\graphicspath{{/}}
\usepackage{lipsum} 
\usepackage[]{siunitx} 
\usepackage{tabularx}
\usepackage[sc]{mathpazo} 
\usepackage[T1]{fontenc} 
\usepackage{microtype} 
\usepackage{booktabs} 
\usepackage{hyperref} 
\usepackage{lettrine} 
\usepackage{paralist} 
\usepackage{transparent} 
\usepackage{titlesec} 
\renewcommand\thesection{\Roman{section}} 
\renewcommand\thesubsection{\Roman{subsection}} 
\titleformat{\section}[block]{\large\scshape\centering}{\thesection.}{1em}{} 
\titleformat{\subsection}[block]{\large}{\thesubsection.}{1em}{} 
\usepackage{listings}

\usepackage{graphicx,import}
\usepackage{amsmath,amsfonts,bm}
\usepackage[ngerman]{varioref,cleveref}

\usepackage{courier}
\usepackage[caption=false]{subfig} 

\usepackage{placeins}
\usepackage{verbatim} 
\begin{document}
\title{
Capturing the shear and secondary compression wave: High frame rate ultrasound imaging in saturated foams
} 

\author{Aichele J.}
\affiliation{Laboratory of Therapeutic Applications of Ultrasound, INSERM \& University of Lyon, Lyon, France}
\author{Giammarinaro B.}
\affiliation{Laboratory of Therapeutic Applications of Ultrasound, INSERM \& University of Lyon, Lyon, France}
\author{Reinwald M.}
\affiliation{Department of Biomedical Engineering, School of Biomedical Engineering \& Imaging Sciences, King's College London, London, UK}
\author{Le Moign G.}
\affiliation{CREATIS Medical Imaging Research Center \& University of Lyon, Lyon, France \& GAUS, University of Sherbrooke, Sherbrooke, Canada}
\author{Catheline S.}
\affiliation{Laboratory of Therapeutic Applications of Ultrasound, INSERM \& University of Lyon, Lyon, France}

\date{\today}


\begin{abstract}
We experimentally observe the shear and secondary compression wave inside soft porous water-saturated melamine foams by high frame rate ultrasound imaging. Both wave speeds are supported by the weak frame of the foam. The first and second compression waves show opposite polarity, as predicted by Biot theory. Our experiments have direct implications for medical imaging: Melamine foams exhibit a similar microstructure as lung tissue. In the future, combined shear wave and slow compression wave imaging might provide new means of distinguishing malignant and healthy pulmonary tissue.

\end{abstract}

\maketitle

\begin{figure}[]

  \subfloat{
	\includegraphics[trim=2.1cm 0cm 3cm 1.5cm,clip,width=0.48\textwidth]{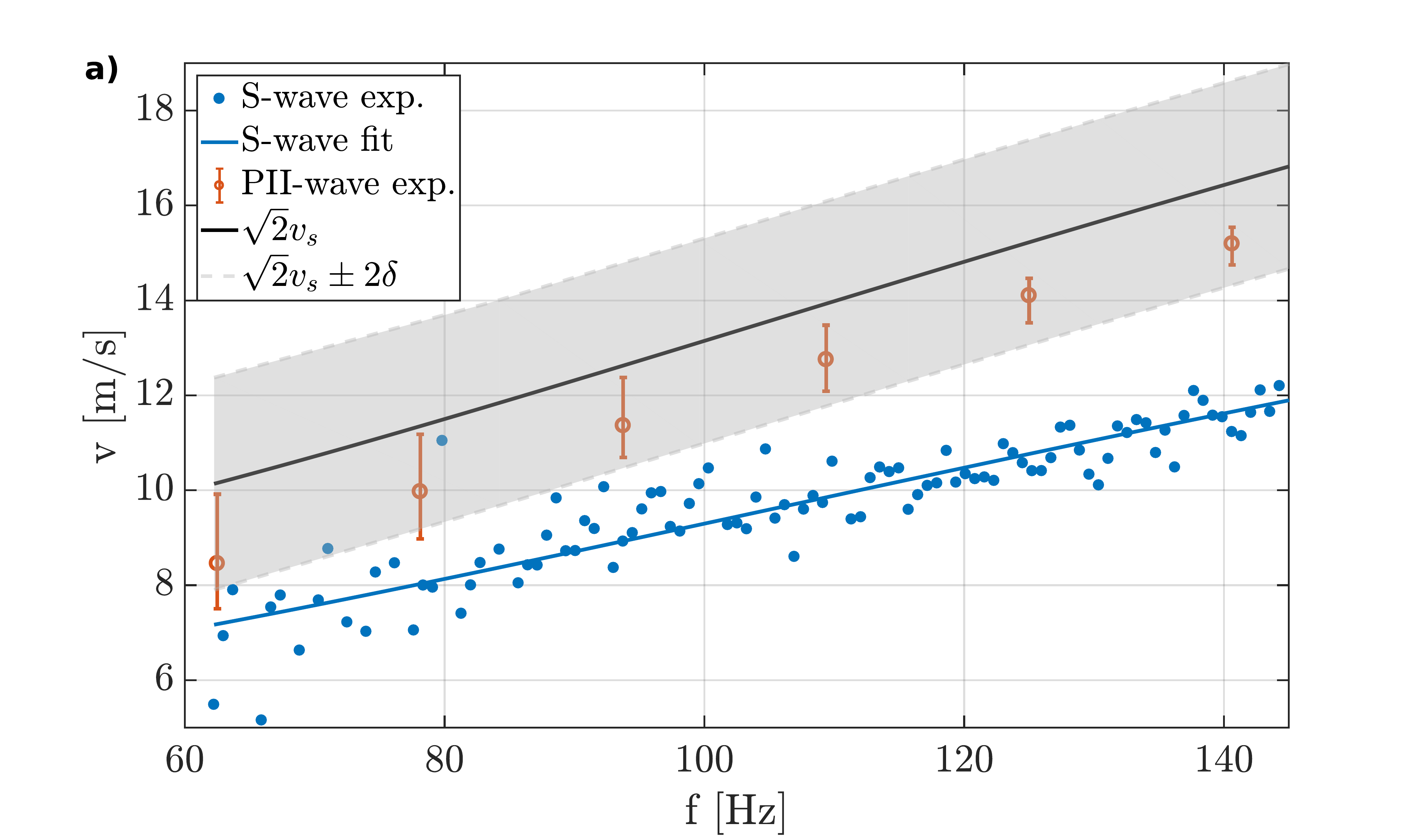}
     }
	\label{fig:DispersionA}
	\subfloat{
	\includegraphics[trim=2.1cm 0cm 3cm 0.5cm,clip,width=0.48\textwidth]{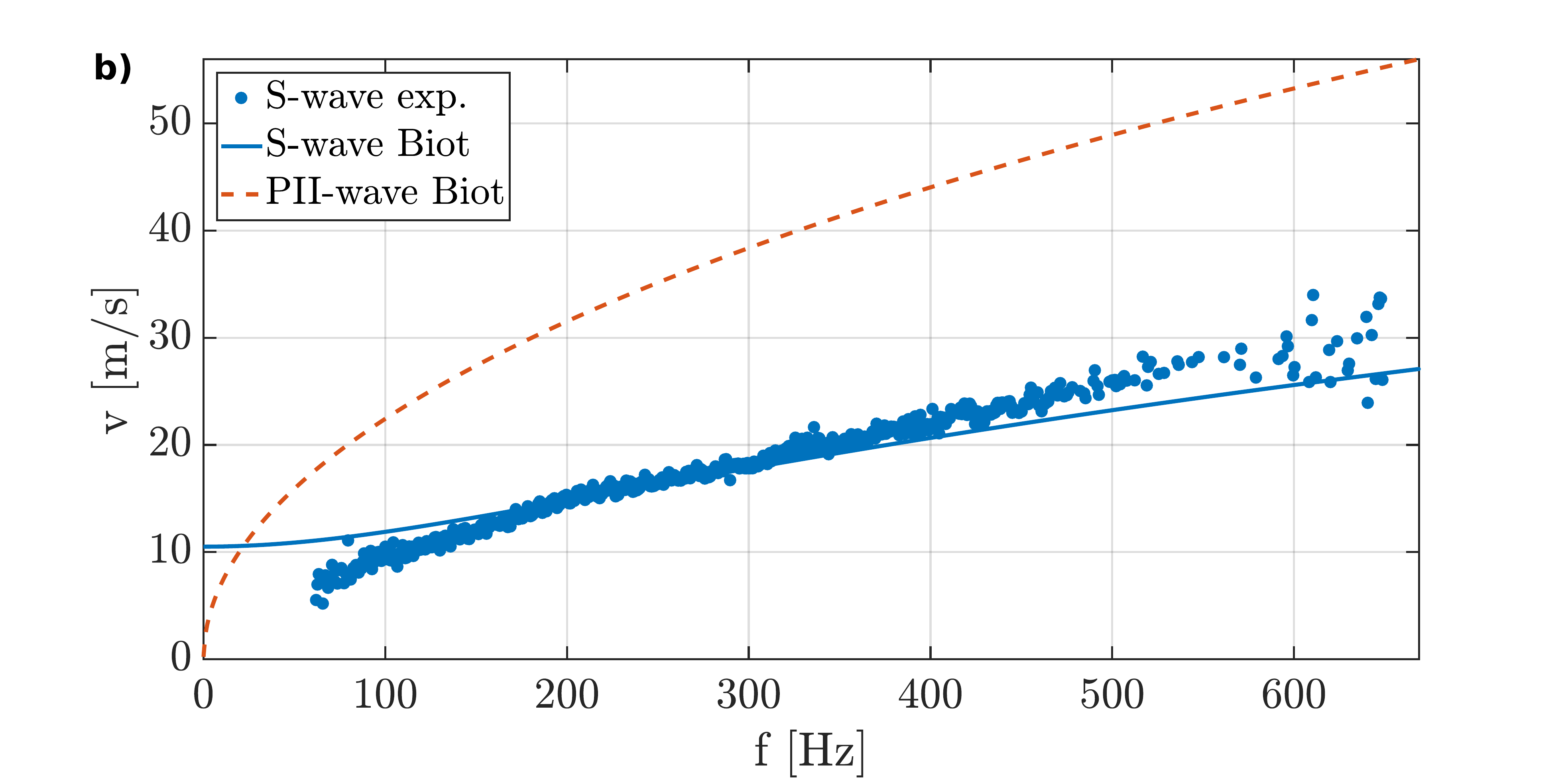}
	}
		\label{fig:DispersionB}
\caption{Experimental and theoretical dispersion. a) Experimental S-wave (blue dots) and PII-wave (red circles) speeds. A sixth-order polynomial fit (blue solid line) of the full frequency band and its \SI{95}{\percent} confidence interval give $v_s$ for Equation \ref{equ:simpleModel}. The resulting PII-wave speeds (black solid line) and its \SI{95}{\percent} confidence interval (gray zone) are displayed. The PII-wave results are the average of three experiments with the maximum deviation indicated by the error bars. b) Experimental S-wave (blue dots) and analytic Biot S- (blue line) and PII-wave (red dashed line).}
\label{fig:Dispersion}
\end{figure}
\begin{figure}[h!]
\includegraphics{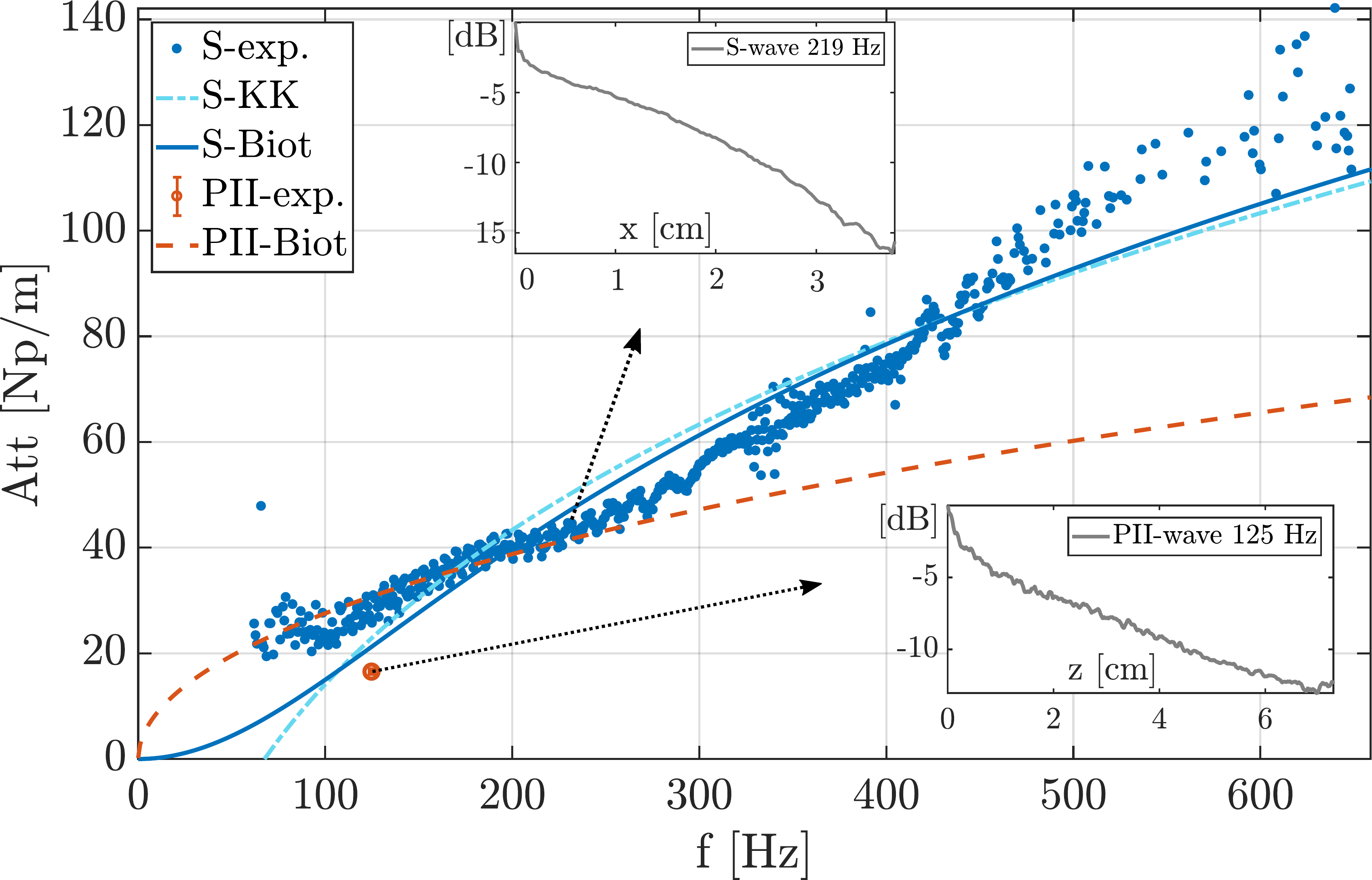}
\caption{\label{fig:s_att_KK} Attenuation measurements, K-K and Biot predictions. The insets show the exemplary amplitude decrease of the S-wave (left) and PII-wave (right). Dots are attenuation measurements, solid blue and dashed red lines are the Biot predictions and the dashed blue line is the S-wave K-K prediction for Equation \ref{equ:PowerLaw}: $\alpha (\omega) = \textit{-119+21}*\omega^{0.29}\quad [\textit{\SI{}{\neper\per\meter}}]$. }
\end{figure}

\paragraph*{}The characterization of wave propagation in porous materials has a wide range of applications in various fields at different scales. In contrast to classical elastic materials, poroelastic materials support three types of elastic waves and exhibit a distinctive dispersion in the presence of viscous fluids \cite{Biot1956a,Plona1980,Boeckx2004}. The first thorough theoretical description of poroelasticity that included dispersion was developed by M. Biot \cite{Biot1956a,Biot1956b}. He predicted a secondary compressional wave (PII-wave), which is often named Biot slow wave. His theory was soon applied in geophysics at large scale for hydrocarbon exploration \cite{GeerstmaPoroMec}. It was later extended to laboratory scale for bone and lung characterization through numerical modeling and medical imaging \cite{Allard,Johnson1987,Champoux1991,Fellah2008,Mizuno2009,Dai2013}.
While poroelastic models have been used to characterize materials and fabrics such as textiles \cite{Alvarez-Arenas2008}, anisotropic composites \cite{Castagnede1998}, snow \cite{Gerling2017} and sound absorbing materials \cite{Boeckx2005a}, experimental detection of the PII-wave remains scarce \cite{Plona1980,Smeulders2001}. In medical imaging, the characterization of the porous lung surface wave has only recently been emphasized \cite{Dai2014,Nguyen2013,Zhang2016,Zhou2018}.
Experimental detection of poroelastic waves is difficult due to their strong attenuation and the diffuse PII-wave behaviour below a critical frequency \cite{Biot1956a,Yang2016,Smeulders2001}.
We overcome this challenge by using in situ measurements from medical imaging. We apply high frame rate (ultrafast) ultrasound for wave tracking \cite{Sandrin1999}, the technique underlying transient elastography \cite{Sandrin1999,Gennisson2003,Catheline2009a,Gennisson2013}, on saturated, highly porous melamine foams. A very dense grid of virtual receivers is placed inside the sample through correlation of backscattered ultrasound images, reconstructing the particle velocity field of elastic waves. The resolution is thus only determined by the wavelength of the tracking ultrasound waves, which is several orders of magnitude lower than the wavelength of the tracked low-frequency waves \cite{Catheline2004,Gennisson2013}. 
Two factors allow us thusly to visualize S- and PII-wave propagation and measure phase speed and attenuation, which, to the best of our knowledge, has not been done before.
Firstly, simple scattering of ultrasound at the foam matrix ensures the reflection image. Secondly, the imaged elastic waves propagate several times slower ($<$ \SI{40}{\metre\per\second}) than the ultrasonic waves ($\approx$ \SI{1500}{\metre\per\second}). 
The measured low-frequency speeds are in agreement with a first approximation that views the foam as a biphasic elastic medium. To take solid-fluid coupling into account, we compare the measured speeds and attenuations with the analytic results of Biot's theory. The S-wave results show a good quantitative prediction, while the PII-wave speeds show a qualitative agreement. Melamine foams have already been used to simulate the acousto-elastic properties of pulmonary tissue due to their common highly porous, soft structure \cite{Mohanty2016, Boeckx2004,Zhou2018}. We thus postulate that our results have possible future implications for lung characterization by ultrasound imaging.
\begin{figure}[!h]
\includegraphics[scale=2]{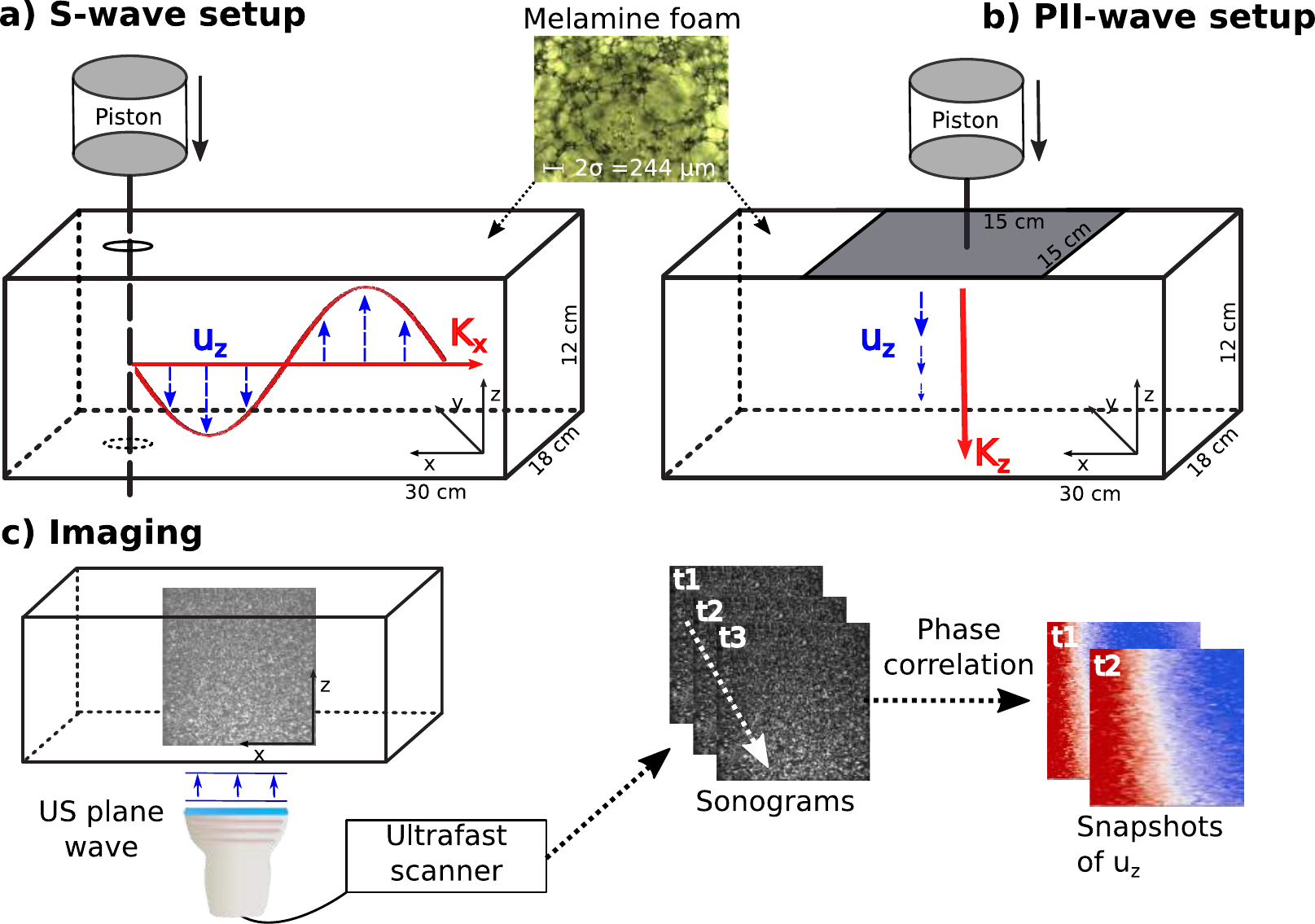}
\caption{Schematic experimental setups of the S-wave (a) and PII-wave excitation (b). Blue arrows signify the polarization of particle motion and red arrows the direction of wave propagation. At the center top, a microscopic photo of the investigated melamine foam is displayed. c) Ultrafast ultrasound imaging principle: The particle velocity maps are retrieved through correlation of subsequent ultrasound images.}
\label{fig:expSetup}
\end{figure}
\paragraph*{}We use a rectangular Basotect\textsuperscript{\textregistered} melamine resin foam of dimensions $x= \SI{30}{\centi\meter}, y= \SI{18}{\centi\metre}, z = \SI{12}{\centi\meter}$, which is fully immersed in water to ensure complete saturation. The foam exhibits a porosity between \SI{96.7} and \SI{99.7}{\percent}, a tortuosity between 1 and 1.02, a permeability between \SI{1.28e-9} and \SI{2.85e-9}{\square\metre} and a density of \SI{8.8}{\kilogram\per\cubic\metre} $\pm$ \SI{1}{\kilogram\per\cubic\metre}. The viscous length $\sigma$ is between \SI{11.24e-5} and \SI{13.02e-5}{\metre}, as indicated in the microscopic photo at the top of Fig. \ref{fig:expSetup}. The foam parameters were independantly measured using the acoustic impedance tube method \cite{Niskanen2017} and a Johnson-Champoux-Allard-Lafarge model \cite{Johnson1987,Champoux1991,Lafarge1997,Niskanen2017}. These measurements serve as input to the analytic Biot model.
Figure \ref{fig:expSetup} shows the setups for the S-wave (a) and PII-wave (b) experiments. A piston (ModalShop Inc. K2004E01), displayed at the top, excites the waves. In a), a rigid metal rod which is pierced through the sponge ensures rod-foam coupling, as well as transverse polarization ($u_z$) of the wave. Excitation is achieved in two ways. Firstly, through a pulse and secondly, through a frequency sweep from 60 to \SI{650}{\hertz}. The excited S-wave propagates along x-dimension ($k_x$) exposing a slight angle due to imperfect alignment of the rod. In b), a rigid plastic plate at the end of a rod excites the compression wave on the upper foam surface and ensures longitudinal polarization. The induced vibration is a Heaviside step function. We undertake three experiments with different excitation amplitudes. In both setups, particle and rod motion are along the z-dimension.
The imaging device is a 128-element L7-4 (Philips) ultrasound probe centered at \SI{5}{\mega\hertz}. Fig. \ref{fig:expSetup}c) schematically indicates the probe position below the foam and the z-polarization of the ultrasonic waves. The probe is connected to an ultrafast ultrasound scanner (Verasonics Vantage{\texttrademark}) which works at 3000 (S-wave) and 2000 (PII-wave) frames per second. Each frame is obtained through emission of plane waves as in \cite{Sandrin1999} and beamforming of the backscattered signals.
In order to visualize the wave propagation, we apply phase-based motion estimation \cite{Pinton2005} on subsequent ultrasound frames. Similar to Doppler ultrasound techniques, the retrieved phase difference gives the relative displacement on the micrometer scale.
Due to the finite size of our sample, reflections from the opposite boundary can occur. Therefore, we apply a directional filter \cite{Buttkus2000} in the $k_x-k_y-f$ domain of the full 2D wavefield \cite{Deffieux2011}. For setup \ref{fig:expSetup}a) (S-wave) the filter effect is negligible, but for setup \ref{fig:expSetup}b) (P-wave), a reflection at the boundary opposite of the excitation plate is attenuated (see Ref. \cite{Suppmat1} and \cite{Suppmat2}). The resulting relative displacement for setup \ref{fig:expSetup}b) is a superposition of the primary compression wave (PI) and the PII-wave. Thus, we additionally apply a spatial gradient in z-direction to isolate the PII-wave displacement.
\begin{figure}[h!]
\includegraphics{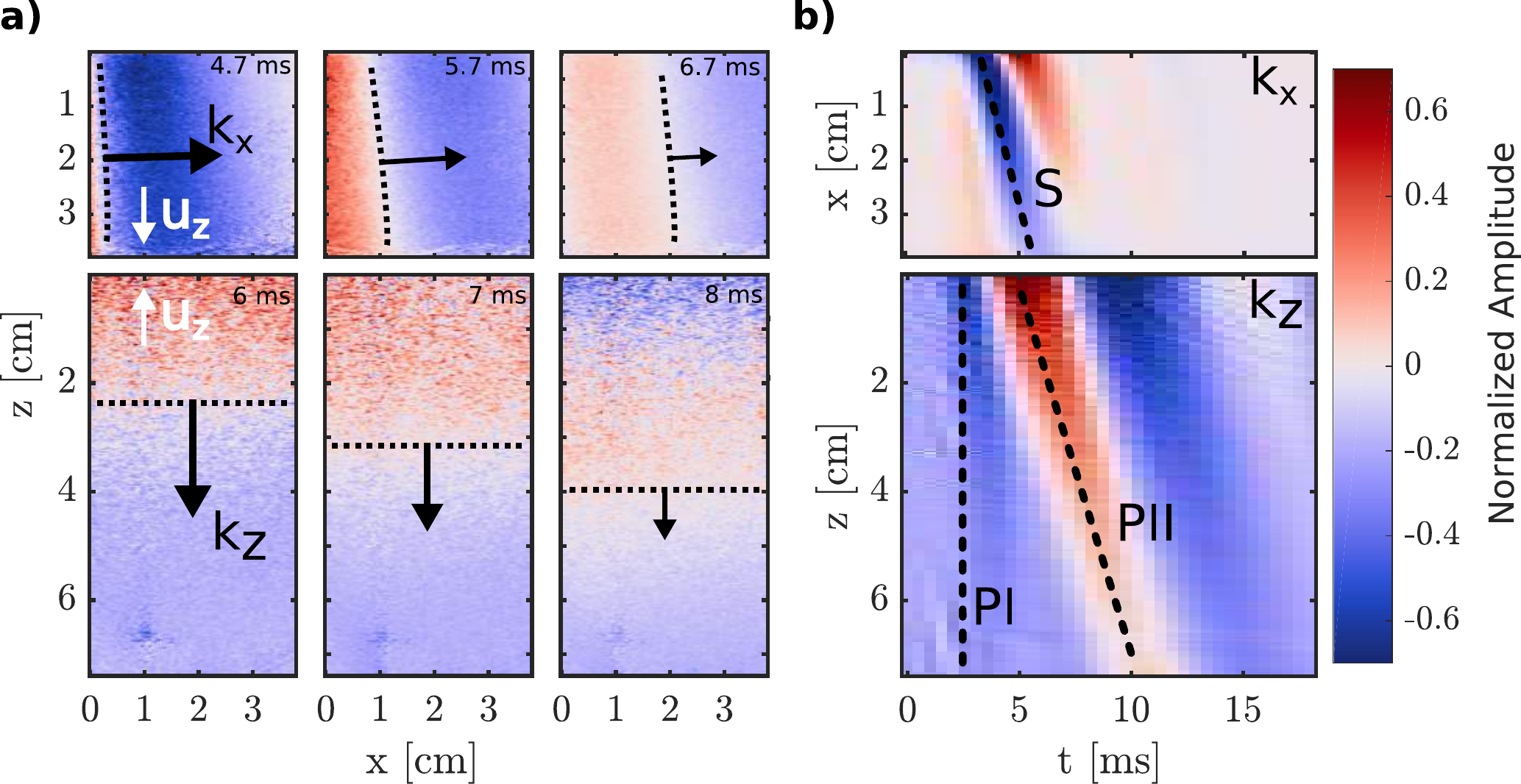}
\caption{\label{fig:Snapshots} Experimental wave-fields for the setups in Fig. \ref{fig:expSetup}. a) Snapshots at three time-steps of a propagating S-wave pulse (top) and PII-wave step (bottom). The top row shows the particle velocity and the bottom its z-gradient. b) Corresponding time-space representation by summation orthogonal to x (top) and z (bottom). In the top row, the S-wave (S) and in the bottom row, the first (PI) and secondary compression wave (PII) can be identified. $u_z$ - Direction of particle motion. $k_{x/z}$ - Direction of wave propagation. The displacement films are in Ref. \cite{Suppmat1} and the unfiltered PII-wave snapshots in Ref. \cite{Suppmat2}.}
\end{figure}
\paragraph*{}Three displacement snapshots of the S- and PII-wave are shown in Fig. \ref{fig:Snapshots}a). The top row is an example of the wave propagation induced by shear excitation as schematically shown in Fig. \ref{fig:expSetup}a). The blue color signifies particle motion $u_z$ towards the probe. A comparison of the wave-fields shows that the plane wave front propagates in the positive x-direction ($k_x$). The bottom row displays the PII-wave for 6, 7 and \SI{8}{\milli\second}. It is excited at the top and propagates with decreasing amplitude in positive z-direction ($k_z$). A summation along the z-dimension for the transverse setup, and along the x-dimension for the longitudinal setup, result in the space-time representations of Fig. \ref{fig:Snapshots}b). They show, that the S- and PII-wave are propagating over the whole length at near constant speed. The PII-wave (PII) is well separated and of opposite polarity from the direct arrival (PI) at \SI{2.5}{\milli\second}. A time-of-flight measurement through slope fitting gives a group velocity of \SI{14.7}{\meter\per\second} (S-wave) and \SI{14.4}{\meter\per\second} (PII-wave). The central frequency is approximately \SI{220}{\hertz} for the S-wave, and \SI{120}{\hertz} for the PII-wave. These values suggest that both speed are governed by the low elastic modulus of the foam. The simplest porous foam model is an uncoupled biphasic medium with a weak frame supporting the S- and PII-wave. In this case, the PI-wave is supported by an in-compressible fluid, which circulates freely through the open pores. The porous compressibility is that of the foam matrix, and its first Lam\'{e} parameter $\lambda_0$ is very small compared to the shear modulus $\mu_0$. Hence, the compression wave speeds $v_{p1,2}$ become:
\begin{align}
\begin{cases}
v_{p1} = \sqrt{\frac{\lambda_f} {\rho_f} } \quad &;\lambda_f >> \mu_f \\
v_{p2} = \sqrt{\frac{(\lambda_0 + 2\mu_0)} {\rho_0}} \approx \sqrt{\frac{2\mu_0} {\rho_0}} \approx \sqrt{2}v_{s}  \quad &;\lambda_0 << \mu_0 
 \end{cases}
\label{equ:simpleModel}
\end{align}
with $\lambda_0$, $\mu_0$ and $\lambda_f$, $\mu_f$ being the first and second Lam\'{e} parameters of the drained sponge and the fluid. $\rho_0 = \rho_{mineral}(1-\phi)$ is the density of the the drained sponge, $\rho_f$ the fluid density, $v_s$ the S-wave speed and $\phi$ is porosity.
This approximation is in accordance with Ref. \cite{Tanaka1973} who investigated the quasi-static behavior of hydrogels.
To assess the dispersion of the observed waves, we apply a fast Fourier transformation and recover the phase velocity and attenuation from the imaginary and real part of the complex signal. For the S-wave, we use a frequency sweep. For the PII-wave, reflections from the boundaries and mode conversions prohibit the exploitation of a chirp, hence we use the Heaviside displayed in Fig. \ref{fig:Snapshots}. Its \SI{-10}{\decibel} bandwidth is limited from 50 to \SI{150}{\hertz}. The phase velocity is directly deduced from a linear fit of the phase value along the propagation dimension. We use a Ransac algorithm \cite{Torr2000} and display values with a $R^2$ of 0.98 and minimum \SI{70}{\percent} inliers. The speed measurements of the S- and PII-wave in their shared frequency band are displayed in Fig. \ref{fig:Dispersion}a). Both curves are monotonously increasing with frequency. To verify Equation \ref{equ:simpleModel} we use a sixth-order polynomial fit (blue line) and its \SI{95}{\percent} confidence interval as input data. The resulting PII-wave speeds (black line) and its \SI{95}{\percent} confidence interval (gray zone) show that the PII-wave experimental data lie within the prediction of Equation \ref{equ:simpleModel}, with a ratio of approximately $\sqrt{2}$ with the S-wave speeds. Figure \ref{fig:Dispersion}b) shows the entire frequency range of the measured S-wave speeds.
\paragraph*{}The elastic model for Equation \ref{equ:simpleModel} cannot account for viscous dissipation. Consequently, we compare the measured dispersion with a second approach, the Biot theory \cite{Biot1956a}. This theory uses continuum mechanics to model a solid matrix saturated by a viscous fluid. The Biot dispersion and PII-wave result from the coupling of fluid and solid displacement \cite{Biot1956a,Biot1956b,Biot1962,Carcione2001,Vogelaar2009,Morency2008,Mavko2009,Allard}.
One drawback is that the theory requires nine parameters. We reduced the degrees of freedom to two by fixing the porous parameters to the values measured by the acoustic impedance tube method in air: $\phi=\SI{0.99}{\percent}$, $\alpha_{\infty} = \SI{1.02}{}$, $k_0=\SI{12.76e-10}{\square\metre}$, $\rho = \SI{8.8}{\kilogram\per\cubic\metre} \pm \SI{1}{\kilogram\per\cubic\metre}$, $BW = \phi$ (Biot-Willis coefficient), and the fluid parameters to literature values for water: $E_{fl} = \SI{2.15e9}{\kilo\pascal}$ (fluid Young's modulus) and fluid viscosity $\eta_f = \SI{1.3e-3}{\pascal\second}$. We optimize the two remaining elastic parameters within literature values of 0.276 to 0.44 for Poisson's ratio and \SI{30} to \SI{400}{\kilo\pascal} for Young's modulus \cite{Geebelen2007,Allard,Deverge2008,Ogam2011,Boeckx2005a}.  To avoid local minima, we first run a parameter sweep in the literature bounds and use the best fit as input to an unconstrained least squares optimization, described in \citep{Nocedal2006}. The resulting Poisson's ratio is 0.39 and the Young's modulus \SI{303}{\kilo\pascal}. However, it should be noted that the optimization is not very sensitive to the Poisson's ratio. For example, a \SI{10}{\percent} increase in Poisson's ratio, results in a $R^2$ of 0.998 between the optimal S-wave solution and the deviation. However, the PII-wave is sensitive to the Poisson's ratio with a $R^2$ of 0.783. In contrast, both waves exhibit a similar sensitivity with a $R^2$ of 0.975 (S-wave) and 0.979 (PII-wave) for a \SI{10}{\percent} increase in Young's modulus. For a detailed sensitivity analysis see Ref. \cite{Suppmat3}. 
The analytic S-wave curve resulting from the minimization is displayed in Fig. \ref{fig:Dispersion}b). It shows good agreement ($R^2 = 0.808$) with the experimental values between 120 and \SI{600}{\hertz}. Below this frequency, wave guiding, present if the wavelength exceeds the dimension of particle motion \cite{Royer1996}, and not taken into account by Biot's infinite medium, might lower the measured speeds. Biot's model overestimates the experimental PII-speeds, but exhibits the same trend.
Furthermore, it predicts that the PII-displacements of the solid and fluid constituent are of opposite sign while they are phase locked for the PI-wave \cite{Biot1956a, Geerits1996}. This leads the PI and PII arrivals to be out-of-phase \cite{Nagakawa2001,Bouzidi2009}, which is confirmed by the time-space representation of Fig. \ref{fig:Snapshots}b). Ultrasound imaging measures the solid displacement only, hence the displacement of PI (blue) and PII (red) is out-of-phase. The phase opposition \cite{Nagakawa2001} and the measured positive PII-dispersion are strong arguments to exclude the presence of a bar wave (S$_0$-mode). 
It should be pointed out, that there is a crucial difference between previous interpretations of the PII-wave \cite{Plona1980,Nagakawa2001,Bouzidi2009} and this study. In geophysics and bone characterization, the PII-wave travels close to the fluid sound speed and the PI-wave close to the sound speed of the rigid skeleton. In contrast to that, our results indicate that the PII-wave speed is governed by the weak frame of the foam and the PI-wave propagates at the speed of sound in water. An equivalent interpretation was given by Ref. \cite{Smeulders2001} for experiments in porous granular media \cite{Paterson1956}.
\paragraph*{}To verify the dispersion results we compare them to attenuation, which for plane waves is described by \cite{Catheline2004}:
\begin{equation}
			A(x+\Delta x)=A(x)e^{{-\alpha (\omega )\Delta x}}
			\label{equ:attenuation}
\end{equation}
where $\omega$ is angular frequency, $\alpha(\omega)$ is attenuation coefficient, $A$ is amplitude and $x$ is measurement direction.
The top left inset in Fig. \ref{fig:s_att_KK} shows the logarithmic amplitude decrease with distance at one frequency of the S-wave. The bottom right inset shows the decrease at the central frequency of the PII-wave. The difference between these experimental curves and the expected linear decrease reflects the difficulties to conduct attenuation measurement by ultrafast ultrasound imaging \cite{Catheline2004}.
We apply a logarithmic fit of the amplitude with distance to retrieve the attenuation coefficient at each frequency, using the RANSAC algorithm described earlier.
The resulting attenuation, displayed in Fig. \ref{fig:s_att_KK}, monotonously increases with frequency.
\paragraph*{}Attenuation and velocity of plane waves can be related through the bidirectional Kramers-Kronig (K-K) relations. They relate the real and imaginary part of any complex causal response function \cite{Toll1956}, which we use to verify our experimental results.
While the original relations are integral functions that require a signal of infinite bandwidth, Ref. \cite{ODonnel,Waters2003,Waters2005} developed a derivative form that is applicable on bandlimited data and has previously been applied by Ref. \cite{Urban2010} on S-wave dispersion. Following Ref. \cite{Catheline2004,Holm2014b}, the attenuation in complex media is observed to follow a frequency power law:
\begin{equation}
\alpha(\omega) =  \alpha_0 + \alpha_1\ \omega^y
\label{equ:PowerLaw}
\end{equation}
and can be related to velocity by \cite{Waters2003,Waters2005}:
\begin{align}
\frac{1}{c(\omega)} - \frac{1}{c(\omega_0)} = \nonumber \qquad \qquad \qquad \qquad \qquad \; \; \; \; \;\;\;  \\*
\begin{cases}
\alpha_1 \tan(\frac{\pi}{2}y)(\omega^{y-1} - \omega_0^{y-1})    &\text{;} \; 0 \leq y \leq 2 \; \\*
\frac{-2}{\pi} \alpha_1 ln \frac{\omega}{\omega_0}  &\text{;} \; y = 1  \;
 \end{cases}
\label{equ:vel_diff}
\end{align}
where $\omega_0$ is a reference frequency, $\alpha_1$ and $y$ are fitting parameters	 and $\alpha_0$ is an offset, typically observed in soft tissues \cite{Szabo2000,Urban2010}.
Since velocity measurements by ultrafast ultrasound imaging are less error prone than attenuation measurements \cite{Catheline2004}, we use Equation \ref{equ:vel_diff} to predict attenuation from velocity. A least squares fit gives the exponent $y$ and the attenuation constant $\alpha_1$ that minimizes Equation \ref{equ:vel_diff} for different reference frequencies. The resulting attenuation model is $\alpha (\omega) = 21*\omega^{0.29}\quad [{\SI{}{\neper\per\meter}}]$, with a reference frequency of \SI{413}{\hertz} and a $R^2$ larger 0.98 for frequencies between \SI{120}{\hertz} and \SI{650}{\hertz}.
The attenuation exponent $y = 0.29$ is an expected value for S-waves in biological tissues \cite{Nasholm2013,Holm2014b}. The K-K relations do not take into account the offset $\alpha_0$. It is introduced by minimizing the least squares of Equation \ref{equ:PowerLaw} and the attenuation measurements. The resulting attenuation curve with $\alpha_0 = \SI{-119}{\neper\per\metre}$ is displayed in Fig. \ref{fig:s_att_KK}. It shows a significant agreement with the attenuation measurements ($R^2=0.9016$) and Biot predictions ($R^2=0.9274$). The successful K-K prediction implies that guided waves have little influence on our measurements above \SI{120}{\hertz}.
The remaining misfit might stem from out-of-plane particle motion, which can introduce an error on amplitude measures by ultrafast ultrasound imaging \cite{Catheline2004}. Furthermore, the low-frequency elastic wave is imaged in the near-field, where it does not show a power law amplitude decrease due to the coupling of transverse and rotational particle motion \cite{Sandrin2004}.
The good agreement between the experimental and theoretical S-wave dispersion and attenuation indicates that the observed S-wave attenuation is due to the interaction between the solid and the viscous fluid.
The PII-wave attenuation of the three experiments converges only at the central frequency of \SI{120}{\hertz} ($\approx$ \SI{16}{\neper\per\meter}). Below this frequency, the measurements are taken on less than two wavelengths of wave propagation and consequently, the exponential amplitude decrease cannot be ensured \cite{Sandrin2004}. A reason for the failure of the Biot theory to quantitatively predict the observed PII-wave dispersion could be viscoelasticity and anisotropy of the foam matrix itself \cite{Deverge2008,Melon1998}. The Biot theory and the model of Equation \ref{equ:simpleModel} have different implications at low frequencies. The Biot PII-wave disappears, whereas it persists as a decoupled frame-borne wave in Equation \ref{equ:simpleModel}. Our velocity measurements support the decoupling hypothesis, but measurements at lower frequencies would be needed to make a definite statement. 

\sisetup{detect-all = true}
\sisetup{detect-none = true}

\paragraph*{}In conclusion, we have showed the first direct observation of elastic wave propagation inside a poroelastic medium. The recorded compression wave of the second kind (PII-wave) propagates at $\sqrt{2}$ times the shear wave speed and is of opposite polarization compared to the first compression wave (PI-wave).
Finally, the measured shear wave dispersion (S-wave) and attenuation are closely related to the fluid viscosity.
These results might have important consequences in medical physics for characterizing porous organs such as the lung or the liver.
\begin{acknowledgments}
We are grateful to Aroune Duclos and Jean-Philippe Groby of the university of Le Mans for introducing us to the world of foams and for measuring the acoustic parameters of the Melamine foam. The project has received funding from the European Union's Horizon 2020 research and innovation programme under the Marie Sklodowska-Curie grant agreement No 641943 (ITN WAVES).
\end{acknowledgments}

%
\end{document}